\documentclass{aa}
\usepackage[varg]{txfonts}
\usepackage{caption}
\usepackage{hyperref}
\usepackage{graphicx}  
\usepackage{float}  
\usepackage{subfigure}  

\usepackage{amsmath}
\usepackage{amssymb}
\usepackage{array}
\usepackage{natbib}
\bibpunct{(}{)}{;}{a}{}{,} 
\hypersetup{
colorlinks=true,
linkcolor=black
}
\usepackage{xcolor}
\usepackage{color}

\usepackage[ruled,vlined]{algorithm2e}
\usepackage{booktabs}
\usepackage{multirow}
\usepackage{graphicx}
\usepackage{txfonts}
\begin{document}
\bibliographystyle{aa} 
   \title{Li-rich and super Li-rich giants produced by element diffusion\thanks{The full version of Table \ref{tab1} only available in electronic form at the CDS via anonymous ftp to cdsarc.u-strasbg.fr (130.79.128.5)}}

\author{Jun Gao
         \inst{1,2}
          \and
          Chunhua Zhu\inst{1,2}
          \and
          Jinlong Yu\inst{3}
          \and
          Helei Liu\inst{1,2}
          \and
          Xizhen Lu\inst{1,2}
         \and
          Jianrong Shi\inst{4,5}
          \and
          Guoliang L\"{u}\inst{1,2*} 
          }

   \institute{School of Physical Science and Technology,
               Xinjiang University, Urumqi 830046, PR China\\
              \email{guolianglv@xao.ac.cn }
         \and Center for Theoretical Physics, Xinjiang University, Urumqi 830046, PR China\\
         \and College of Mechanical and Electronic Engineering, Tarim University, Alar 843300, PR China\\
         \and CAS Key Laboratory of Optical Astronomy, National Astronomical Observatories, Chinese Academy of Sciences, Beijing 100101, China\\
         \and School of Astronomy and Space Science, University of Chinese Academy of Sciences, Beijing 100101, China\\
             }



\abstract
{About 0.2-2\% of giant stars are Li-rich, whose lithium abundance\,(A(Li)) is higher than 1.5\,dex. Among them, near 6\% are super Li-rich with A(Li) exceeding 3.2\,dex. Meanwhile, the formation mechanism of these Li-rich and super Li-rich giants is still under debate.}
{Considering the compact He core of red giants, attention is paid to the effect of element diffusion on A(Li). 
	In particular, when the He core flash occurs, the element diffusion makes the thermohaline mixing zone extend inward and connect to the inner convection region of stars. 
	Then, a large amount of $^7$Be produced by the He flash can be transferred to stellar surface, finally turning into $^7$Li. 
	Thus, the goal of this work is to propose the mechanism of A(Li) enrichment and achieve the consistency between the theoretical and observation data.}
{Using the Modules for Experiments in Stellar Astrophysics\,(MESA), we simulate the evolution of low-mass stars, with considering the effects of element diffusion on the Li abundances.  
	The timescale ratio of Li-rich giants to normal giants is estimated by population synthesis method. 
	Then we get the theoretical value of A(Li) and make a comparison with observations. }
{Considering the influence of element diffusion in the model results in the increase of lithium abundance up to about 1.8\,dex, which can reveal Li-rich giants. 
	Simultaneously, introducing high constant diffusive mixing coefficients\,($D_{\rm mix}$) with the values from $10^{11}$ to $10^{15}$cm$^2$ s$^{-1}$ in the model allows A(Li) to increase from 2.4 to 4.5\,dex, which can explain the most of Li-rich and super Li-rich giant stars. The population synthesis method reveals that the amount of Li-rich giants among giants is about 0.2-2\%, which is consistent with observation estimated levels.
}
{In our model,  the element diffusion mainly triggered by the gravity field changes the mean molecular weight at the junction zone between the stellar envelope and the He core, which  makes the thermohaline mixing region expanding to the inner convection region of stars. 
A transport channel, efficiently transporting $^7$Be in the hydrogen burning region of the star to the convective envelope where $^7$Be decays into $^7$Li, is formed. 
Combing a high constant diffusive mixing coefficients, the transport channel can explain the origin of Li-rich and super Li-rich giant, even  the most super Li-rich giants.}


\maketitle
%

\section{Introduction}
Lithium\,(Li) is one of the important elements to study the origin of the universe. 
In the evolution of low-mass stars, Li begins to deplete in the main sequence\,(MS) stage. 
This process experienced the first dredge-up and some deep mixing, and most of the Li will be consumed.
The phenomenon has been predicted by a standard staller evolution model\citep{Deepak2019}, and confirmed by numerous observations \citep{Brown1989,Lind2009,Liu2014,Kirby2016}.

Since the first giant star with a high Li abundance was discovered by \cite{Wallerstein1982}, the development of the related research field challenged the traditional stellar evolution mechanism. 
These stars were called Li-rich giants, with a classic definition of A(Li)\,$\geq$\,1.5\,dex\citep{Iben1967a,Iben1967b}.
\footnote{Here, A(Li) is the Li abundance expressed as A(Li) = log\,[n\,(Li)\,/\,n\,(H)]\,+\,12, where n\,(Li) and n\,(H) is the number density of Li and hydrogen, respectively.} 
It is noteworthy that Li-rich giants are extremely rare, accounting about 1\%-2\% or even less among all normal giants\citep{Brown1989,Kumar2011,Liu2014,Casey2016,Kirby2016,Monaco2011,Li2018,Gao2019}. 
In particular, the ratios estimated
from some large survey programs are $\sim$0.9\% from Gaia-ESO
survey \citep{Casey2016, Smiljanic2018}, $\sim$0.8\%
from RAVE \citep{Ruchti2011} and $\sim$0.2\%-0.3\%
from SDSS and GALAH data \citep{Martell2013,Deepak2019}.  

In the past 40 years, various Li-rich giant stars have been successively identified. 
Based on the Large Sky Area Multi-Object Fiber Spectroscopic Telescope (LAMOST) low resolution spectra acquired in China: in 2018, a star with the highest Li abundance was found,with A(Li) $\sim$ 4.5\,dex\citep{Yan2018}.
The Li-rich giants with A(Li) higher than 3.2\,dex are called super Li-rich giants, whose amount among all Li-rich giants is about 6\%\citep{Singh2021}. 
In the period from October 2011 to June 2019, a total of 10,535 Li-rich giants with A(Li)\,$\geq$\,1.5\,dex were screened from the LAMOST low-resolution spectra, which allowed one to expand the existing observation sample database by about 5 times and greatly enriches research samples.

There are two main hypotheses about the origin of Li-rich giant stars: one is that Li comes from the outside of stars, such as planetary engulfment or pollution of binary companion stars \citep{Stephan2018,Lodders2019,Holanda2020}. 
The other is through $^{3}$He($\alpha$,$\gamma$)$^{7}$Be(e,$\nu$)$^{7}$Li, also known as  Cameron-Fowler\,(CF) mechanism \citep{Cameron1955,Cameron1971}. 
In the latter process, $^{7}$Be produced at a high-temperature region of H burning has to be quickly carried away to a low-temperature region, such as the convective envelope, where it will decay into $^{7}$Li, and this fresh $^{7}$Li will survive. 
As is predicted by \cite{Schwab2020} that there is enough $^{7}$Be in the H-burning shell prior to the first helium\,(He) subflash.

Combining the Kepler and spectroscopy information, \cite{Yan2021} confirmed that most Li-rich stars are red clump\,(RC), while a few of them are red giant branch\,(RGB). 
In order to explain the Li enhancement in RC stars, \cite{Mori2020} introduces a neutrino magnetic moment\,(NMM), which shows that $^{7}$Be production becomes more active owing to the fact that the delay of He flash makes themohaline mixing more effective when the NMM is excited.
Thus Li was produced at the tip of the red giant branch\,(TRGB). 
However, they did not attempt to explain the origin of super Li-rich giants. 
At the same time, Kumar has also used the GALAH DR2\citep{Buder2018} and Gaia DR2\citep{Gaia2018} data to confirm that the Li abundance of RC stars is 40 times higher than those at TRGB\citep{Kumar2020}. 
Perhaps, this abnormality of Li abundance might be due to a complex interaction between TRGB and RC phase. 
The most significant event between these two stages is the occurrence of a He flash, especially the first, strongest He subflash.

Recently, \cite{Schwab2020} has reported that the He flash induced mixing links the H-burning shell and the convective zone, and plenty of $^{7}$Be circulate toward the cooler convective zone to turn into $^{7}$Li through the CF mechanism. 
The model proposed by \cite{Schwab2020} implies there is the possibility of very high Li abundances\,(see Fig.4), although the author notes that these high Li abundances are quickly depleted in their model. 
Therefore, searching for a physical mechanism of Li enrichment\,(including super Li-rich giant stars) to obtain the consistency between observation and theory is of great significance for fundamental and applied research. 
It is well known that the surface chemical abundance of a star can be affected by many factors such as convection, thermohaline mixing or element diffusion\citep[e. g.,][]{Kippenhahn1980,Dupuis1992,Zhu2021}. 
The effects of convection and thermohaline mixing on Li abundances of RGB stars have been investigated by \cite{Yan2018} and \cite{Martell2021}. 
However, the element diffusion is seldom considered. 
Element diffusion is a dynamic process that changes the distribution of chemical elements in stars. 
It is mainly the result of the joint action of pressure, temperature, material concentration, and other factors. 
Element diffusion plays a very important role in the stellar evolution, especially in the chemical element distribution on the stellar surface\citep{Semenova2020}.

In this paper, we consider the effects of the element diffusion impact on the Li abundance for stars in the RC phase. 
In particular, Sect. 2 describes the details of the stellar model and element diffusion. 
Sect. 3 presents the Li abundances predicted by our model and the comparative analysis of the observation and theoretical results. 
The summary is given in Sect. 4.  
\section{Models}
We use Modules for Experiments in Stellar Astrophysics\,(MESA, [rev.\,12778]; \cite{Paxton2011,Paxton2013,Paxton2015,Paxton2018,Paxton2019})  to construct one-dimensional low-mass stellar models.
MESA adopts the equation of state of \cite{Rogers2002} and \cite{Timmes2000} and the opacity of \cite{Iglesias1996,Iglesias1993} and \cite{Ferguson2005}.

Our model use the standard MESA $pp\_and\_cno\_extras$ nuclear network, which includes 25 species and the reactions covering the pp-chains and CNO-cycles.
We adopt nuclear reaction rates compiled by JINA REACLIB \citep{Cyburt2010}.
Treatment of electron screening is based on \cite{Alastuey1978} and \cite{Itoh1979}.
The mass loss formula in \cite{Reimers1975} is adopted.
We use the electron-capture rate on $^{7}$Be from \cite{Simonucci2013}, as made available in machine-readable form by \cite{Vescovi2019}.
Models are initialized on the pre-main sequence with the \cite{Asplund2009} solar abundance pattern and Z = 0.014. This initializes Li to the meteoritic abundance A(Li) = 3.26\,dex.

The size of convective zone depends on mixing length parameter. We adopt a mixing length of 1.8 times the pressure scale height. These models include thermohaline mixing($\alpha_{\rm th}$) using the \cite{Kippenhahn1980} prescription with an efficiency of $\alpha_{\rm th}$ = 100. This gives the deep mixing necessary to destroy $^{7}$Li on the first ascent giant branch.

Elemental diffusion in stars is mainly driven by a combination of pressure gradients\,(or gravity), temperature gradients, compositional gradients, and radiation pressures.
The main driving factor of element diffusion is gravity sedimentation\citep{Paxton2015,Paxton2018}. In the model, we input a mixing diffusive coefficient $D_{\rm mix}$$\sim$($\Delta$R)$^{2}$/($\Delta$t) to show the efficiency of element diffusion in different regions.
Standard stellar evolution theory points out that the H-burning core in the MS stage continuously generates helium elements, which are deposited into the stellar interior, so as to form a helium core. The He core reaches a certain mass, it will begin to burn.
We speculate that the influence of element diffusion will affect the element abundance on the star surface, which will affect the formation process of helium core, the occurrence time of the first He flash, and then stimulate the thermohline mixing to make the mixing process more sufficient.

By solving the Burgers equation \citep{Burgers1969}, \cite{Thoul1994} proposed a general method to arrange the whole set of equations into a single matrix equation, that is to input Burgers equation into the matrix structure without readjusting any number. There is no approximation of the relative concentrations of various species, nor is there any limitation on the number of elements to be considered. Therefore, this method is suitable for a wide variety of astrophysical problems. Using the method of \cite{Thoul1994}, MESA can calculate the diffusion of chemical elements in stellar interior \citep{Paxton2015,Paxton2018}.

The inputs provided by the MESA model are the number densities $n_{\rm s}$, temperature $T$, the gradients of these quantities $d$\,ln\,$n_{\rm s}$/$dr$ and $d$\,ln\,$T$/$dr$, species mass in atomic units $A_{\rm s}$, species mean charge as an average ionization state $\overline{Z}_{\rm s}$, and the resistance coefficients $K_{\rm st}$, $z_{\rm st}$, $z_{\rm st}$$^{\prime}$ and $z_{\rm st}$$^{\prime}$$^{\prime}$, as defined by Equation (86) in \cite{Paxton2015}. In our model, diffusion coefficients $D_{\rm mix}$ derived from \cite{Paquette1986} and updated by \cite{Stanton2016}.
By calculating the characteristic duration of the first He flash and the characteristic length scale of the mixing region, this suggest a effective mixing diffusion coefficient requires $D_{\rm mix}$\,\textgreater\,10$^{10}$\,cm $^2$ s$^{-1}$\citep{Schwab2020}. Together with the mean ionization states, these are key parts of the input physics that determine the diffusion of all ions. The additional acceleration terms $g_{\rm rad,s}$ of radiation suspension is set to zero by default.
\section{Result}
\begin{figure}
	\resizebox{\hsize}{!}{\includegraphics{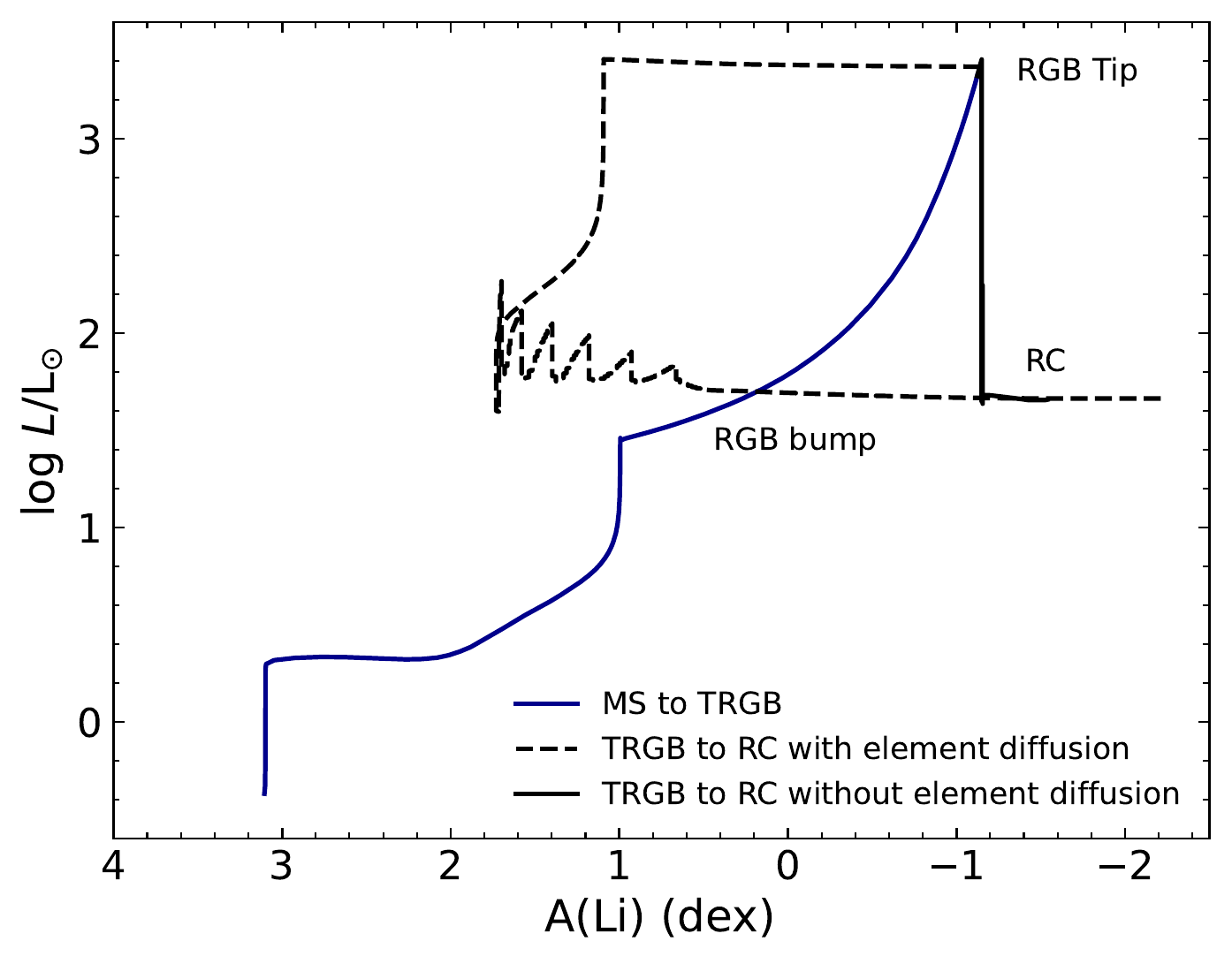}}
	\caption{The evolution of Li on the stellar surface for 1\,$M_\odot$ model. The x axis is the Li abundance. The y axis represents the logarithm of luminosity. The lines indicates the evolution process of Li abundance.}
	\label{1}
\end{figure}
According to \cite{Yan2021}, the Li-rich RGB and RC stars have different mass distributions with the peaks at about 1.7\,$M_\odot$ and 1.2\,$M_\odot$, respectively.
In general, when the stellar mass at a zero age main sequence\,(ZAMS) is lower than 0.9\,$M_\odot$, the star hardly evolves into a giant phase. Simultaneously, when the stellar mass is larger than 1.8\,$M_\odot$, the temperature in the envelope of the star in the giant phase is sufficiently high. As a result, beryllium\,(Be) elements produced by the H-burning shell are quickly destroyed and cannot be brought to the surface of the star.
Therefore, we calculated the evolution of Li abundance for the models with masses of 0.9, 1.0, 1.2, 1.4, 1.6, and 1.8\,$M_\odot$. 
For simplicity, a 1.0\,$M_\odot$ model was taken as a sample.
\subsection{Effect of element diffusion}
The solid line in Fig. \ref{1} shows the standard evolutionary track.
It shows that low mass star samples usually start from the MS turnoff with A(Li)\,$\sim$\,3.2\,dex and suffer depletion through the first dredge-up. 
Then the star reaches the RGB bump with a luminosity of 10$^{1.6}$\,$L_\odot$, where it begins to consume Li rapidly again\citep{Iben1967a, Charbonnel2007, Lattanzio2015}. 
The black solid line represents the stellar evolution without element diffusion at Stage II. 
It shows the luminosity drops sharply to the level of RC, whilst maintaining the RGB tip A(Li)\citep{Kumar2020}. 
The black dotted line denotes the model with element diffusion. 
A(Li) successfully increases from -1 to 1.8\,dex in Stage II. 
As a result, the element diffusion improves the efficiency of thermohaline mixing, and its activity range gets connected with the convective zone of stellar interior. 
Also, the $^{7}$Be in the H-burning shell get transferred to the convective envelope, where they decay into $^{7}$Li through the CF mechanism. 
This figure reveals that the increase of Li abundance up to 1.8\,dex can be realized by considering the element diffusion.

Almost all explanations proposed for Li-rich giants involve the He flash\citep{Kumar2020,Mori2020,Schwab2020}.
Based on the standard model of stellar structure and evolution, the He-core burning occurs when the mass of the He core increases to $M_{\rm He}$\,$\approx$\,0.45\,$M_\odot$\citep{Thomas1967,Bildsten2012}.
Therefore, the stellar evolution in this work was divided into two stages: Stage\,I is from the MS stage to $M_{\rm He}$\,=\,0.45\,$M_\odot$.
The next range until the RC stage is called Stage\,II.

Figure \ref{1} displays the evolution trajectory of Li abundance on the surface of a 1$M_\odot$ star without and with the element diffusion. 
As shown by the dark-blue solid line, the evolution of Li abundance in both models was similar during Stage I. 
In the MS phase, the Li abundance is kept constant.
When a low-mass star evolves into a red giant, it undergoes the first dredge-up. 
After the first dredge-up, Li starts to rapidly decrease because the Li elements are mixed up the stellar interior material by the envelope deepen process and then are diluted. 
This depletion is due to dilution as the envelope deepens and mixes up interior material heavily depleted in Li. 
Before the RGB tip, the Li abundance continued to drop due to the thermohaline mixing. 
From the ZAMS to the RGB tip, the Li abundance on the stellar surface decreased by about 4 orders of magnitudes, which was consistent with the results of \cite{Kumar2020}.
\begin{figure*}
	\centering
	\includegraphics[width=17cm]{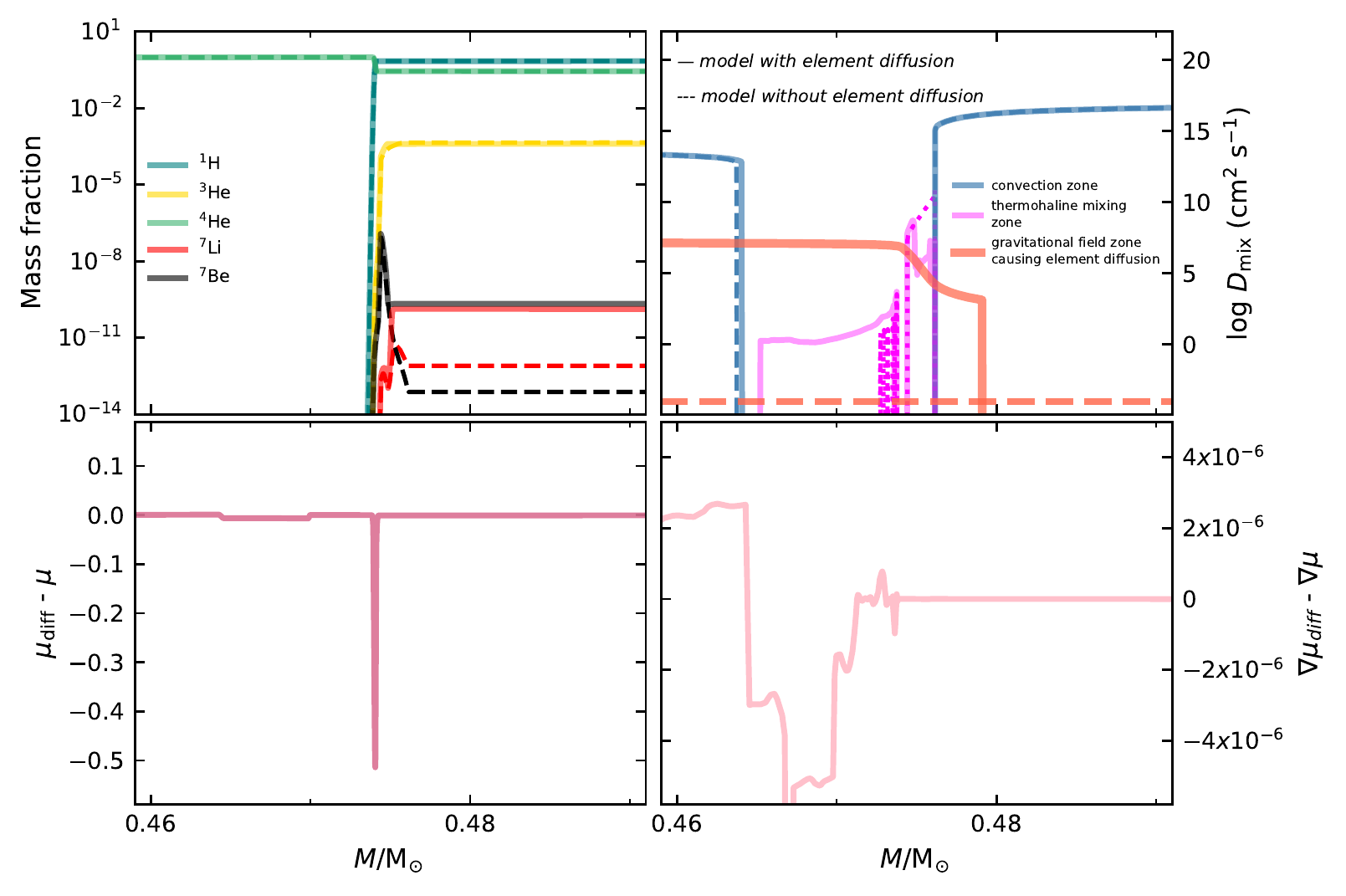}
	\caption{ Profiles of element abundances (top left corner) and diffusion coefficient (top right corner) ($D_{\rm mix}$) on the first He flash.
		The solid lines are for model with element diffusion, and the dashed lines are for models without element diffusion.
		The left and right panels at the bottom show the relative changes between the mean molecular weights ($\mu_{\rm diff}$ and $\mu$) with and without element diffusion, and between the mean molecular weight gradients ($\nabla \mu_{\rm diff}$ and $\nabla \mu$) with and without element diffusion, respectively.}
	\label{2}
\end{figure*}

The He flash emerges on the RGB tip. The Li abundance in the model without element diffusion was falling during Stage II. 
However, A(Li) in the case of element diffusion increased from -1 to 1.8\,dex. It indicates that element diffusion can enhance the Li abundance on the stellar surface. 
The main reason for this phenomenon is depicted in Fig. \ref{2}. Especialy, a compact He core is formed on the RGB tip, whose strong gravity can produce the efficient element diffusion, resulting in the expansion of the thermohaline mixing zone. 
As shown in the top-right panel of Fig. \ref{2}, for the model with element diffusion, the thermohaline mixing zone connected to the convection zone in the stellar interior in which the He flash occurred, that is, it extended to the deeper interior of the hydrogen burning region.
Therefore, the large amounts of $^{7}$Be, produced by the H-burning shell, could be transferred to the stellar surface, finally turning into $^{7}$Li. 
On the other hand, for the model without element diffusion, $^{7}$Be could not or rarely be brought to the stellar envelope due to the disconnection of the stellar interior convection zone and the thermohaline mixing zone.

Thermohaline convection is a turbulent mixing process that can occur in stellar radiative regions whenever the mean molecular weight increases with radius. 
In some cases, it can have a significant observable impact on stellar structure and evolution\citep{Ulrich1972,Kippenhahn1980,Brown2013,Garaud2018}. 
The left and right panels at the bottom of the Fig. \ref{2} show the relative changes of mean molecular weights ($\mu_{\rm diff}$ and $\mu$) and mean molecular weight gradients ($\nabla \mu_{\rm diff}$ and $\nabla \mu$) respectively at the junction zone between the stellar envelope and the He core. 
It can be seen that due to the influence of element diffusion, the $\mu_{\rm diff}$ and $\nabla \mu_{\rm diff}$ have decreased significantly.
The local decrease of mean molecular weight can drive a more efficient thermohaline mixing. 
It expands to the inner convection region of stars, which is shown by pink lines in the top-right panel. 
These indicate that element diffusion can greatly affect the mean molecular weight and the mean molecular
weight gradient, which leads to the expansion of the mixing region of thermohaline convection.

The essence of this phenomenon is that the element diffusion mainly triggered by the gravity field suppress the the mean molecular weight and changes the element concentration gradient at the junction zone between the stellar envelope and the He core. 
It is one of the most important factors which affect the thermohaline mixing. 
It is well known that the mean molecular weight greatly affects the efficiency and range of thermohaline mixing, and make the thermohaline
mixing region expanding to the inner convection region of stars. 
That is, the thermohaline mixing can occur in the internal area of hydrogen burning, and bring products and by-products of nuclear reactions to the surface. 
Thus, a transport channel, efficiently transporting $^7$Be in the hydrogen burning region of the star to the convective envelope where $^7$Be decays into $^7$Li, is formed.

\subsection{Effect of element diffusion with constant diffusive coefficients}
Although A(Li) on the surface of the star predicted by the model with element diffusion can increase up to about 1.8\,dex,
it cannot explain the formation of the super Li-rich giants.

\begin{figure}
	\resizebox{\hsize}{!}{\includegraphics{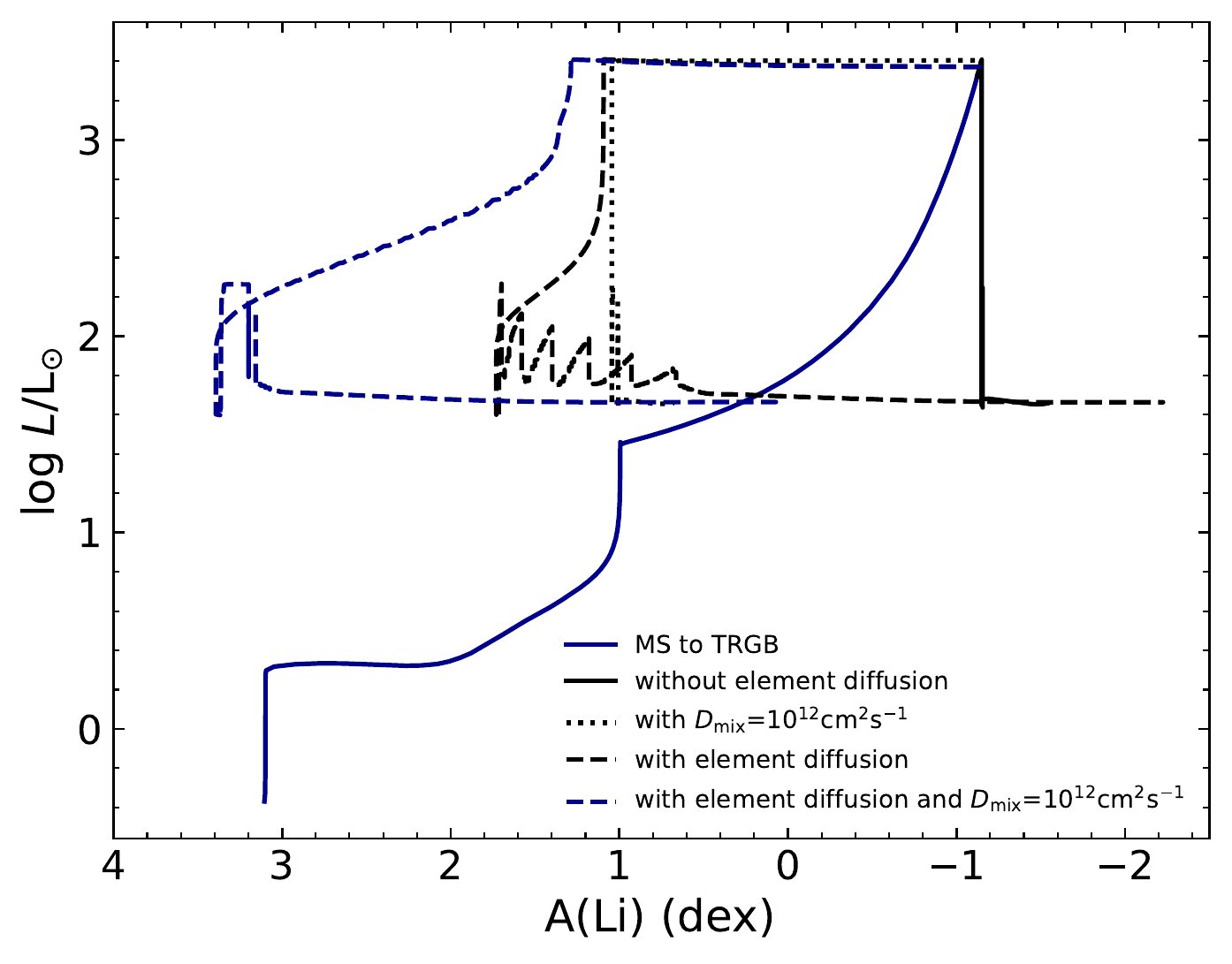}}
	\caption{Similar to Fig. \ref{1}, but for the evolutional tracks of Li abundances of four models.}
	\label{3}
\end{figure}
\begin{figure*}[ht]
	\begin{tabular}{c@{\hspace{3pc}}c}
		\centering
		\includegraphics[width=8.5cm]{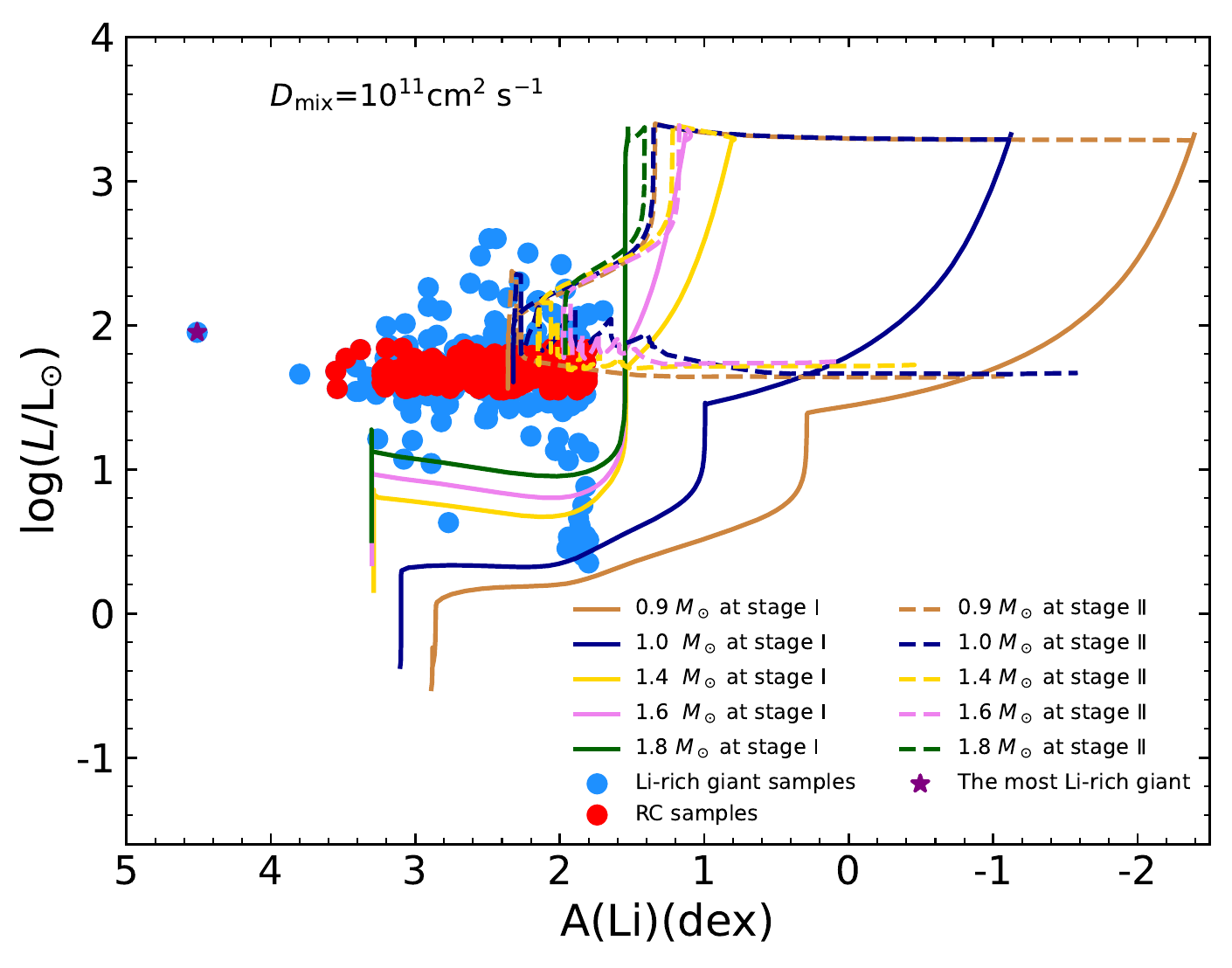}&
		\includegraphics[width=8.5cm]{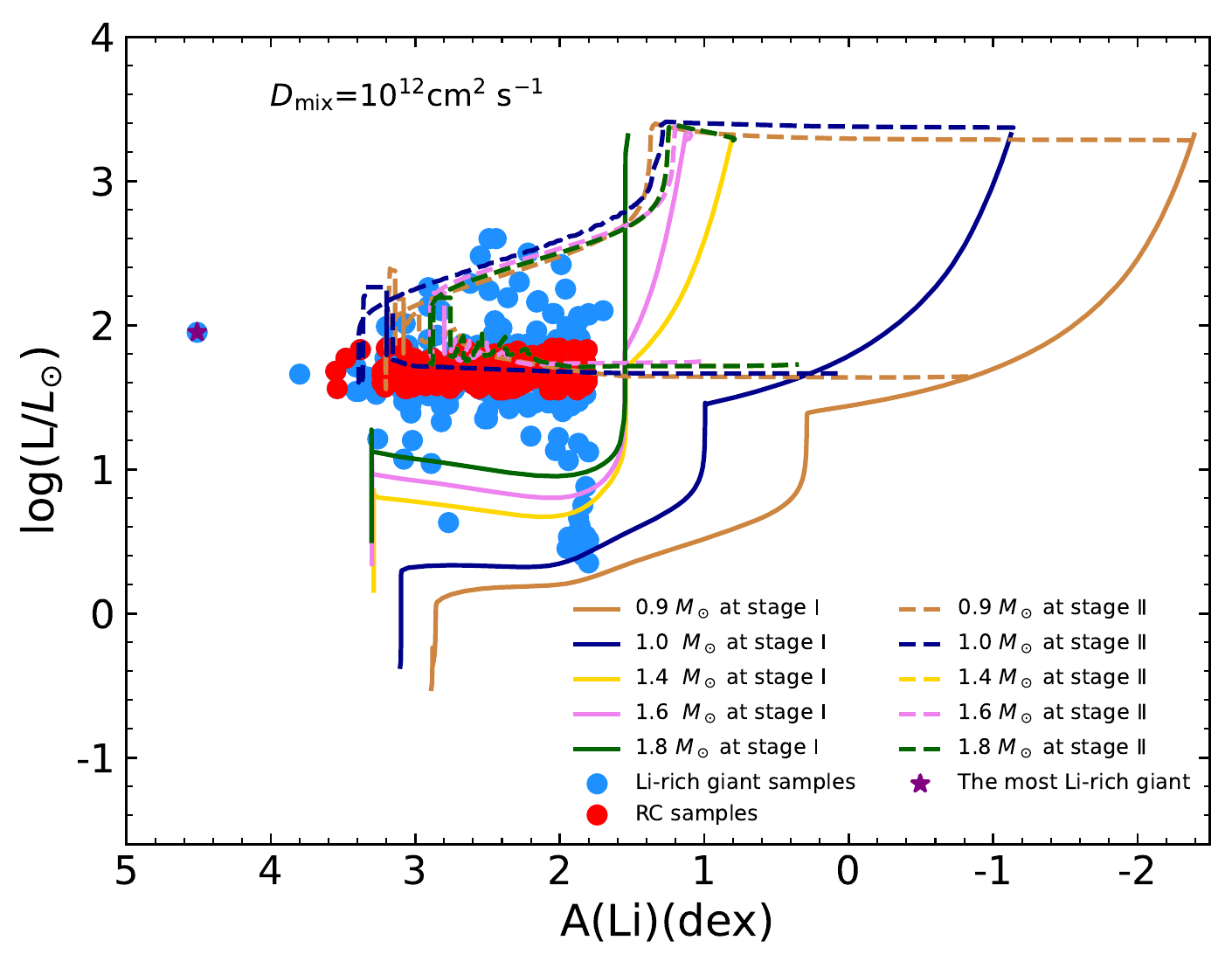}\\
		\includegraphics[width=8.5cm]{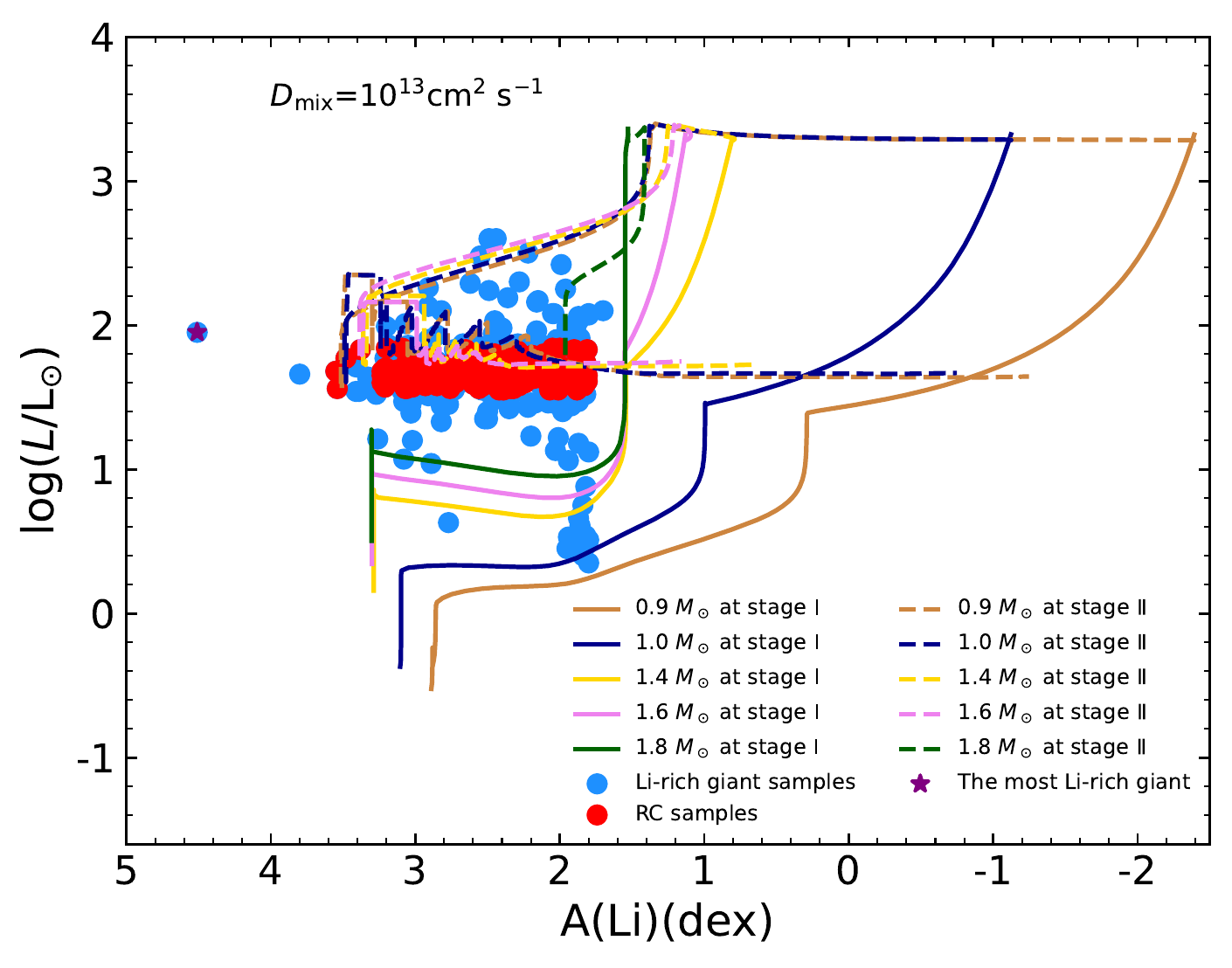}&
		\includegraphics[width=8.5cm]{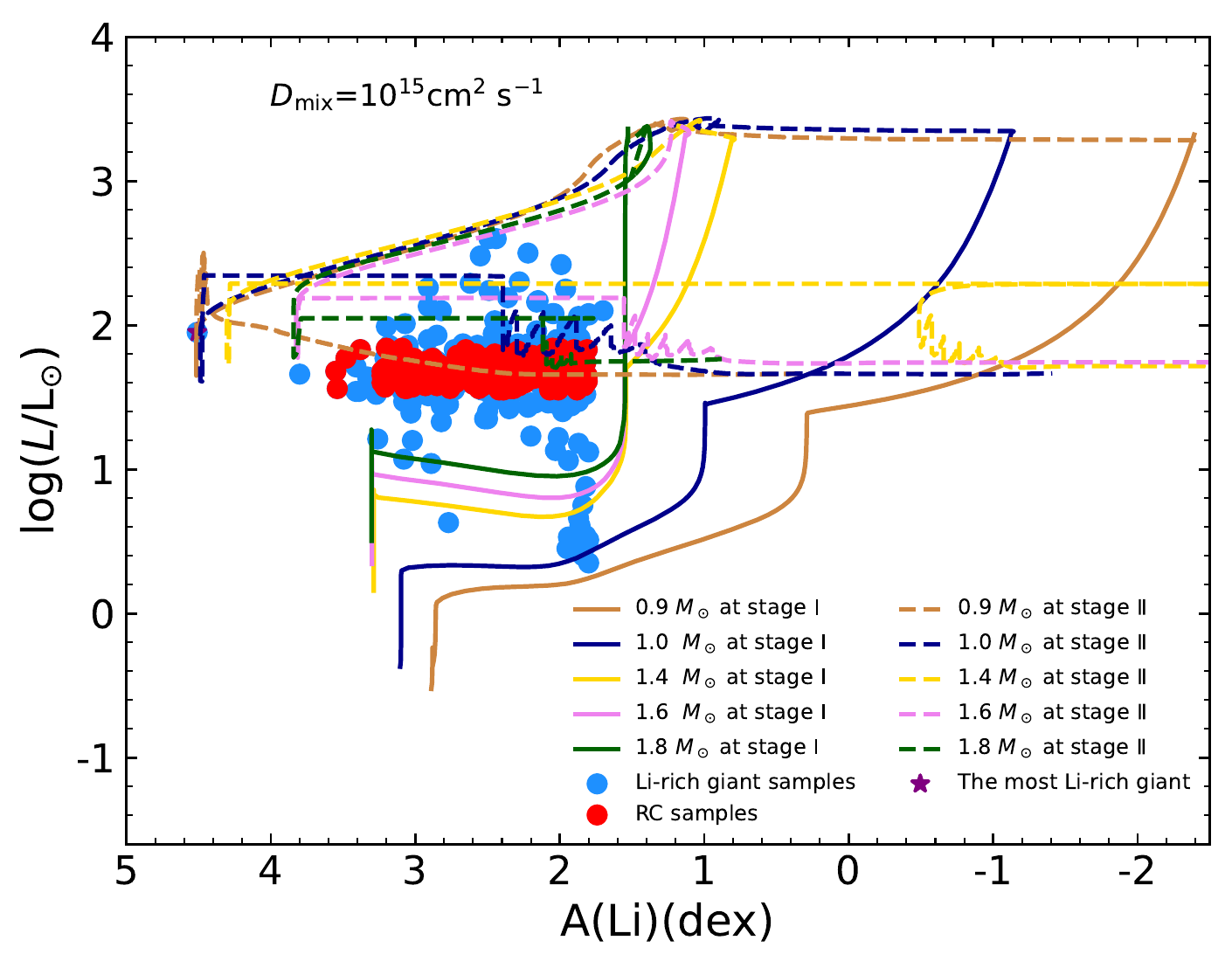}\\
	\end{tabular}
	\caption{Similar to Fiure 1, but for the models with different masses and constant diffusive mixing coefficient of $D_{\rm mix}$ to $10^{11}$ cm$^2$ s$^{-1}$, $10^{12}$ cm$^2$ s$^{-1}$, $10^{13}$ cm$^2$ s$^{-1}$ and $10^{15}$\,cm$^2$ s$^{-1}$.
		Different values are displayed in the upper left corner of each subgraph. The solid and dashed lines show the evolutional tracks at Stages\,I and II, respectively. The red and blue circles are the Li-rich giant stars and RC stars listed in Table 1, respectively.
		The star represent the most Li-rich giant star, TYC\,429-2097-1 observed by \cite{Yan2018}.}
	\label{4}
\end{figure*}
\begin{figure*}
	\centering
	\includegraphics[width=17cm]{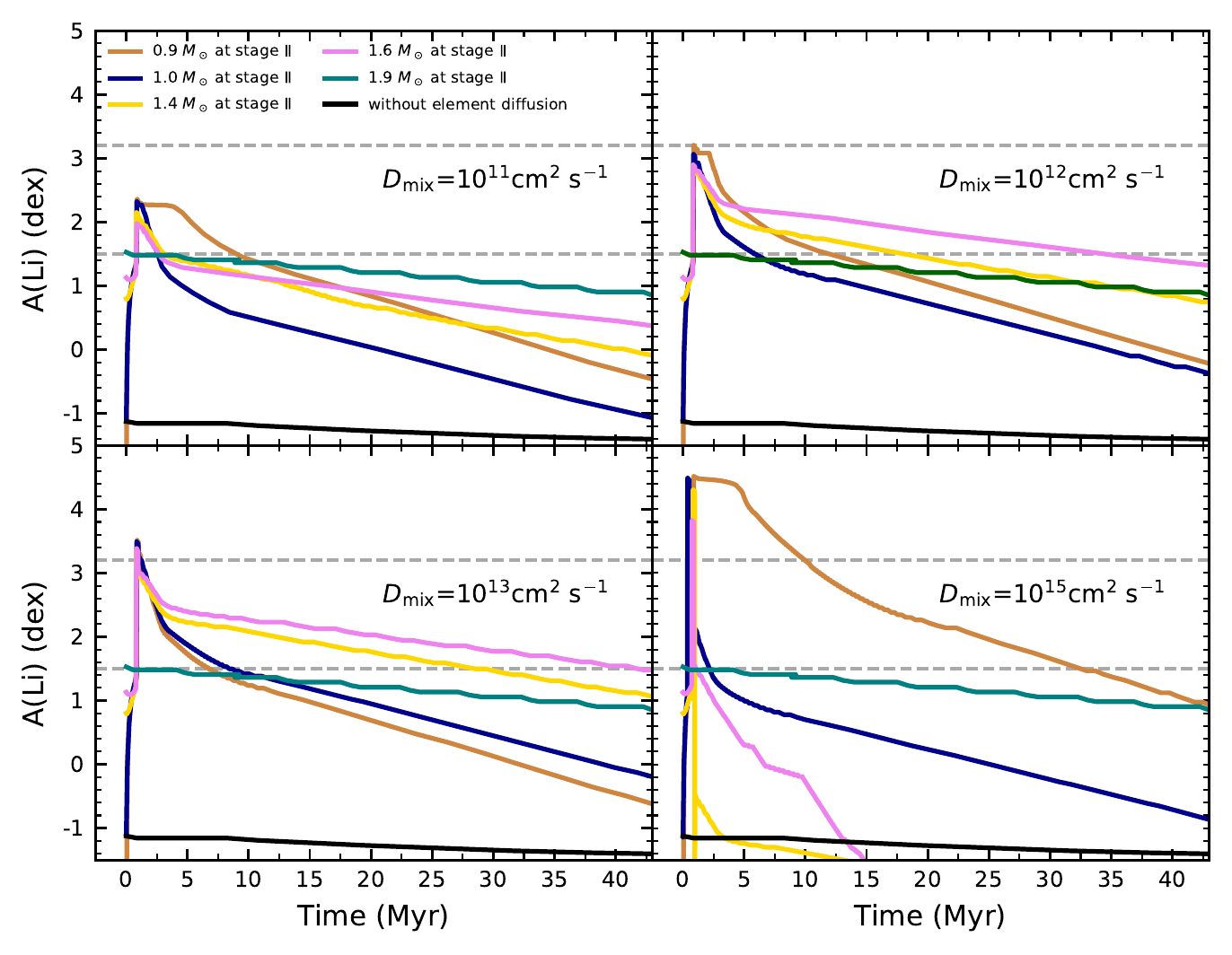}
	\caption{The evolution of A(Li) after the first He flash with time. 
		The solid lines represents the increase produced with element diffusion and different constant diffusive coefficient( $D_{\rm mix}$). The dotted line indicates the A(Li)\,$\sim$\,1.5\,dex and A(Li)\,$\sim$\,3.2\,dex.}
	\label{5}
\end{figure*}
According to Fig. \ref{2}, the diffusive coefficient is a very important factor for the formation of the Li-rich giant. 
Very recently, in order to produce the Li-rich giant, \cite{Schwab2020} has considered that turbulent convective motions can excite internal gravity waves, and a chemical mixing occurs when the luminosity of the He flash\,($L_{\rm He}$) is higher than $10^4$\,L$_\odot$. 
In particular, the effective diffusive mixing coefficient was estimated to be about $10^{11}$\,cm$^2$ s$^{-1}$. 
Given the constant diffusive mixing coefficients $D_{\rm mix}$ of $10^{10}$, $10^{12}$ and $10^{14}$\,cm$^2$ s$^{-1}$ from different models, the maximum value of A(Li) calculated by \cite{Schwab2020} was about 3.6\,dex. 
Meanwhile, the model proposed by \cite{Schwab2020} fails to explain the origin of super Li-rich giants.

In this section, combining the element diffusion and constant diffusive mixing coefficients, we calculate the evolution of A(Li) on the stellar surface. 
Following \cite{Schwab2020}, we assume a constant diffusive mixing coefficient for the mixing when $L_{\rm He}$\,>\,$10^4$\,L$_\odot$. 
According to Fig. \ref{2}, $D_{\rm mix}$ during the He flash could reach $10^{15}$\,cm$^2$ s$^{-1}$. 
Therefore, in this work,  $D_{\rm mix}=10^{11}$, $10^{12}$, $10^{13}$ and $10^{15}$\,cm$^2$\,s$^{-1}$ were adopted in different models.

Figure \ref{3} displays the A(Li) evolution on the surface of a star with the mass of 1\,$M_\odot$. 
Using the model with a constant diffusive mixing coefficient of $10^{12}$\,cm$^2$ s$^{-1}$ but without element diffusion could enhance the Li abundance up to about 1.0\,dex, which agreed with the result of \cite{Schwab2020}. 
In the model with element diffusion, the Li abundance could be increased to 1.8\,dex. 
At this time, the inner convection region was connected with the thermohaline mixing zone to form a channel for transporting $^{7}$Be elements, which greatly increased the $^{7}$Li on the stellar surface. 
Surprisingly, the model combining the element diffusion and the constant diffusive mixing coefficient exhibited an increase of A(Li) up to 3.4\,dex. 
The reason for this is that the diffusive mixing coefficient improves the mixing efficiency of the channel excited by the element diffusion. Therefore, A(Li) in the model with the element diffusion and the constant diffusive mixing coefficient can keep a constant value above 3.0\,dex.
\subsection{Li-rich giants and super Li-rich giants}
In recent years, many large survey programs have revealed the existence of numerous Li-rich giants.
Combining the astrometric data from the Gaia satellite \citep{Gaia2016} with spectroscopic abundance surveys (such as GALAH survey, LAMOST survey), twenty Li-rich abundances could be identified.
Based on GALAH DR2 and DR3 surveys, \cite{Deepak2019}, \cite{Deepak2020}, \cite{Kumar2020} and \cite{Martell2021} measured the Li abundances of 1872 giant stars.
According to LAMOST survey,  \cite{Singh2019} and \cite{Yan2021} explored the Li abundances of 456 giant stars which are in the Kepler field.
In order to compare with the theoretical results with observation samples in this work, we selected 351 published Li-rich giants with precise values of luminosity, temperature and Li abundance as our samples\,(see Table \ref{tab1}).

Figure \ref{4} depicts the observed data on Li-rich giants and the theoretical results for the models with different $D_{\rm mix}$s and masses. In this study, an increase in $D_{\rm mix}$ from $10^{11}$ to $10^{15}$\,cm$^2$\,s$^{-1}$ led to a rise in A(Li) from 2.4 to 4.5\,dex. For the models with element diffusion and $D_{\rm mix}>10^{12}$\,cm$^2$ s$^{-1}$, the evolutionary tracks passed through most of observed samples of super Li-rich giants (A(Li)\,$\geq$\,3.2\,dex), Li-rich giants (A(Li\,$\geq$\,1.5\,dex) to the normal giants. Especially, the value of A(Li), calculated in the model with element diffusion and $D_{\rm mix}=10^{15}$\,cm$^2$ s$^{-1}$, reached 4.5\,dex, which could explain the Li abundance of the most super Li-rich giants.
\subsection{Population synthesis for Li-rich giants}
As mentioned in the Introduction, the Li-rich giants among giants are scarce\,(about 0.2-2\%). Based on the models in this work, we estimate the theoretical ratio by the population synthesis method which is used in the previous investigations by our group\citep{L2006,L2009,L2013,L2020,Yu2019,Yu2021,Zhu2021}.

In the population synthesis method for single-star systems, the initial mass function (IMF) is the most important input parameter.  
The IMF used in the present research was derived from the stellar distribution toward both Galactic poles as well as that within 5.2\,pc of the Sun by \cite{Kroupa1993}.
Based on this IMF, 10$^6$ stars were produced by Monte Carlo calculation. 
In order to estimate their percentage, the lifetimes of Li-rich giants and giants were afterward estimated.

Figure \ref{5} displays the evolution of A(Li) in all models. After the first He flash, the element diffusion firstly initiates and enhances the Li abundance.
According to \cite{Schwab2020},  the constant diffusive mixing coefficient can work when $L_{\rm He}>$ 10$^4$\,$L_\odot$.
The Li abundance greatly increased after about 0.2\,Myr of the first He flash, and remained constant within several Myr due to a constant $D_{\rm mix}$.
Of course, A(Li) decreases when the stellar luminosity is lower than 10$^4$\,$L_\odot$. 
When the mass is greater than about 1.9\,M$_\odot$,  the element diffusion and constant diffusive mixing coefficient are not excited because the temperature within the stellar envelope is too high. 
Simultaneously, the lifetimes of the Li-rich giants decrease with the increase of $D_{\rm mix}$. 
The main reason for this is that the high diffusive coefficient accelerates the circulation process of elements, so that Be or Li elements can be quickly carried to a high-temperature zone and destroyed. 

In this study, MESA was applied to calculate the evolution of stars with initial masses of 0.9\,M$_\odot$, 1.0\,M$_\odot$, 1.2\,M$_\odot$, 1.4\,M$_\odot$, 1.6\,M$_\odot$ and 1.8\,M$_\odot$. 
Through a linear interpolation method, the lifetimes of Li-rich giants and giants for these 10$^6$ stars were estimated. 
Then, the percentage of Li-rich giants among giants was assessed.
In particular, the ratio values of models with $D_{\rm mix}=10^{11}$, $10^{12}$, $10^{13}$ and $10^{15}$\,cm$^{2}$ s$^{-1}$  were 0.5, 1.2, 1.1 and 0.2\%, respectively, which was consistent with the observational estimates.

Very recently, \cite{Zhang2021} has shown that the deterioration of Li in the RC stage is not so obvious and the low ratio\,(such as 0.2\%) may be an anomaly. 
If it is true, the high diffusive coefficient\,(such as $10^{15}$\,cm$^{2}$ s$^{-1}$) may be undesirable. 
It means that the model proposed in this study fails to produce all super Li-rich giants, especially the most Li-rich giant star TYC 429-2097-1. 
Thus, the binary merging model proposed by \cite{Zhang2020} may be competitive.
\section{Conclusions}
Considering the element diffusion, we used MESA to calculate the evolution of Li abundance. 
The element diffusion mainly triggered by the gravity field suppress the the mean molecular weight and changes the element concentration gradient at the junction zone between the stellar envelope and the He core. 
The local decrease of mean molecular weight greatly affects the efficiency and range of thermohaline mixing, and make the thermohaline mixing region expanding to the inner convection region of stars. 
A transport channel, efficiently transporting $^7$Be in the hydrogen burning region of the star to the convective envelope where $^7$Be decays into $^7$Li, is formed. 
Therefore, a large amount of $^7$Be, produced by the He flash, could be transferred to the stellar surface, finally turning into $^7$Li. 
However, the value of A(Li) could be increased up to only 1.8\,dex, which was insufficient to produce super Li-rich giant stars.

In turn, combining the element diffusion and constant diffusive mixing coefficients enabled one to increase the theoretical A(Li) values from 2.4 to 4.5\,dex by increasing $D_{\rm mix}$ from $10^{11}$ to $10^{15}$\,cm$^2$ s$^{-1}$.
This means that the element diffusion in the proposed model can result in the extension of the thermohaline mixing zone and its connection with the stellar interior convection zone.
Then, $^7$Be produced by He burning can be mixed in the stellar envelope. 
The high diffusive mixing coefficient can improve the efficiency of $^7$Be transfer to the stellar surface. 
Therefore, our model can produce the Li-rich giants, and even the most of super Li-rich giants. 
The results provided by population synthesis method were also consistent with the observations, which confirmed the feasibility of this mechanism. 
However, the accuracy of the results in the model under consideration may be affected by the uncertain input parameter, $D_{\rm mix}$. 
Since calculating an accurate $D_{\rm mix}$ is beyond the scope of this work, attention is rather paid to the diffusive mixing coefficients.
\begin{acknowledgements}
We are grateful to anonymous referee for careful reading of the paper and constructive criticism.
This work received the generous support of the National
Natural Science Foundation of China, project Nos. U2031204, 11863005,
12163005, and 12090044, 
the science research grants from the China Manned Space Project with NO. CMS-CSST-2021-A10,
and the Natural Science Foundation of Xinjiang
No.2021D01C075.
\end{acknowledgements}

\begin{table*}[htb]
\centering
\caption{From about 11000 observational samples, the 351 Li-rich giant stars whose luminosities are measured are selected in this work. The observational data come from the references listed in the last column. (The full version of this Table is available at the CDS.)}
\begin{tabular}{ccccccc}
\hline\hline
   Object ID & Teff (K) & log g (dex) & [Fe/H] (dex) & log(L/$L_\odot$) & A(Li)&Reference\\
\hline
HD 8676	    &	4860	&	2.95	&	0.02	&	1.68	&	3.55	&	Kumar et al.(2011)\\
HD 10437	&	4830	&	2.85	&	0.1	    &	1.77	&	3.48	&	Kumar et al.(2011)\\
HD 12203	&	4870	&	2.65	&	-0.27	&	1.69	&	2.08	&	Kumar et al.(2011)\\
HD 37719	&	4650	&	2.4	    &	0.09	&	1.76	&	2.71	&	Kumar et al.(2011)\\
HD 40168	&	4800	&	2.5	    &	0.1	    &	2.1	    &	1.7	    &	Kumar et al.(2011)\\
HD 51367	&	4650	&	2.55	&	0.2 	&	1.59	&	2.6	    &	Kumar et al.(2011)\\
HD 77361	&	4580	&	2.35	&	-0.02	&	1.66	&	3.8	    &	Kumar et al.(2011)\\
HD 88476	&	5100	&	3.1	    &	-0.01	&	1.87	&	2.21	&	Kumar et al.(2011)\\
HD 107484	&	4640	&	2.5	    &	0.18	&	1.78	&	2.14	&	Kumar et al.(2011)\\
HD 118319	&	4700	&	2.2	    &	-0.25	&	1.68	&	2.02	&	Kumar et al.(2011)\\
HD 133086	&	4940	&	2.98	&	0.02	&	1.7	    &	2.14	&	Kumar et al.(2011)\\
HD 145457	&	4850	&	2.75	&	-0.08	&	1.61	&	2.49	&	Kumar et al.(2011)\\
HD 150902	&	4690	&	2.55	&	0.09	&	1.83	&	2.65	&	Kumar et al.(2011)\\
HD 167304	&	4860	&	2.95	&	0.18	&	1.93	&	2.85	&	Kumar et al.(2011)\\
HD 170527	&	4810	&	2.85	&	-0.1	&	1.69	&	3.12	&	Kumar et al.(2011)\\
TYC 429-2097-1	&	4696	&	2.25	&	-0.36	&	1.95	&	4.51	&	Yan et al. (2018)\\
Gaia DR2 6423511482552457344	&	4828.68	&	2.84	&	0.18	&	1.56	&	3.54	&	Deepak and Reddy(2019)\\
Gaia DR2 6216747182780840576	&	4773.08	&	2.69	&	0.12	&	1.54	&	3.41	&	Deepak and Reddy(2019)\\
Gaia DR2 3080569351805501824	&	4995.53	&	2.6	   &	0.03	&	1.71	&	3.41	&	Deepak and Reddy(2019)\\
Gaia DR2 5920543908525756800	&	4815.52	&	2.68	&	0.14	&	1.54	&	3.39	&	Deepak and Reddy(2019)\\
Gaia DR2 5676420200792553600 	&	4854.1	&	2.31	&	-0.11	&	1.83	&	3.38	&	Deepak and Reddy(2019)\\
Gaia DR2 6721793108675117440	&	4911.04	&	2.45	&	-0.04	&	1.64	&	3.33	&	Deepak and Reddy(2019)\\
Gaia DR2 4488063566731544960	&	4778.94	&	2.37	&	-0.02	&	1.52	&	3.27	&	Deepak and Reddy(2019)\\
Gaia DR2 2939800046333110272 	&	4985.2	&	2.56	&	-0.13	&	1.55	&	3.26	&	Deepak and Reddy(2019)\\
Gaia DR2 4168437628181576192 	&	4749.88	&	2.71	&	0.19	&	1.21	&	3.26	&	Deepak and Reddy(2019)\\
Gaia DR2 5229729170925959552	&	5038.38	&	2.79	&	-0.15	&	1.65	&	3.24	&	Deepak and Reddy(2019)\\
Gaia DR2 5293680581122445184 	&	4832.25	&	2.5	   &	-0.01	&	1.65	&	3.23	&	Deepak and Reddy(2019)\\
Gaia DR2 5242382659974594688	&	4786.39	&	2.8	   &	0.28	&	1.6	   &	3.23	&	Deepak and Reddy(2019)\\
Gaia DR2 5628302754467688576 	&	4868.94	&	2.37	&	-0.2	&	1.57	&	3.21	&	Deepak and Reddy(2019)\\
Gaia DR2 3202012502737830784	&	4906.32	&	2.42	&	-0.14	&	1.77	&	3.21	&	Deepak and Reddy(2019)\\
Gaia DR2 5460011229840058880 	&	4813.14	&	2.63	&	0.16	&	1.7	   &	3.21	&	Deepak and Reddy(2019)\\
Gaia DR2 5452473905831060480	&	4711.6	&	2.17	&	-0.3	&	1.84	&	3.2	   &	Deepak and Reddy(2019)\\
Gaia DR2 6162898261508964992	&	4835.99	&	2.58	&	0.01	&	1.7	   &	3.2	   &	Deepak and Reddy(2019)\\
Gaia DR2 6779302244026689920	&	4541.3	&	2.17	&	-0.48	&	1.99	&	3.2	   &	Deepak and Reddy(2019)\\
Gaia DR2 3496188144418768640 	&	4776.14	&	2.59	&	0.04	&	1.63	&	3.2	   &	Deepak and Reddy(2019)\\
...	&	...	&	...	&	...	&	...	&	...	&	...\\
   \hline
 \label{tab1}
\end{tabular}
\end{table*}

\medskip
\bibliography{gjaa} 

\begin{thebibliography}{74}
\expandafter\ifx\csname natexlab\endcsname\relax\def\natexlab#1{#1}\fi

\bibitem[{{Alastuey} \& {Jancovici}(1978)}]{Alastuey1978}
{Alastuey}, A. \& {Jancovici}, B. 1978, \apj, 226, 1034

\bibitem[{{Asplund} {et~al.}(2009){Asplund}, {Grevesse}, {Sauval}, \&
  {Scott}}]{Asplund2009}
{Asplund}, M., {Grevesse}, N., {Sauval}, A.~J., \& {Scott}, P. 2009, \araa, 47,
  481

\bibitem[{{Bildsten} {et~al.}(2012){Bildsten}, {Paxton}, {Moore}, \&
  {Macias}}]{Bildsten2012}
{Bildsten}, L., {Paxton}, B., {Moore}, K., \& {Macias}, P.~J. 2012, \apjl, 744,
  L6

\bibitem[{{Brown} {et~al.}(1989){Brown}, {Sneden}, {Lambert}, \&
  {Dutchover}}]{Brown1989}
{Brown}, J.~A., {Sneden}, C., {Lambert}, D.~L., \& {Dutchover}, Edward, J.
  1989, \apjs, 71, 293

\bibitem[{Brown {et~al.}(2013)Brown, Garaud, \& Stellmach}]{Brown2013}
Brown, J.~M., Garaud, P., \& Stellmach, S. 2013, The Astrophysical Journal,
  768, 34

\bibitem[{{Buder} {et~al.}(2018){Buder}, {Asplund}, {Duong}, {Kos}, {Lind},
  {Ness}, {Sharma}, {Bland-Hawthorn}, {Casey}, {de Silva}, {D'Orazi},
  {Freeman}, {Lewis}, {Lin}, {Martell}, {Schlesinger}, {Simpson}, {Zucker},
  {Zwitter}, {Amarsi}, {Anguiano}, {Carollo}, {Casagrande}, {{\v{C}}otar},
  {Cottrell}, {da Costa}, {Gao}, {Hayden}, {Horner}, {Ireland}, {Kafle},
  {Munari}, {Nataf}, {Nordlander}, {Stello}, {Ting}, {Traven}, {Watson},
  {Wittenmyer}, {Wyse}, {Yong}, {Zinn}, {{\v{Z}}erjal}, \& {Galah
  Collaboration}}]{Buder2018}
{Buder}, S., {Asplund}, M., {Duong}, L., {et~al.} 2018, \mnras, 478, 4513

\bibitem[{{Burgers}(1969)}]{Burgers1969}
{Burgers}, J.~M. 1969, {Flow Equations for Composite Gases}

\bibitem[{{Cameron}(1955)}]{Cameron1955}
{Cameron}, A.~G.~W. 1955, \apj, 121, 144

\bibitem[{{Cameron} \& {Fowler}(1971)}]{Cameron1971}
{Cameron}, A.~G.~W. \& {Fowler}, W.~A. 1971, \apj, 164, 111

\bibitem[{{Casey} {et~al.}(2016){Casey}, {Ruchti}, {Masseron}, {Randich},
  {Gilmore}, {Lind}, {Kennedy}, {Koposov}, {Hourihane}, {Franciosini}, {Lewis},
  {Magrini}, {Morbidelli}, {Sacco}, {Worley}, {Feltzing}, {Jeffries},
  {Vallenari}, {Bensby}, {Bragaglia}, {Flaccomio}, {Francois}, {Korn},
  {Lanzafame}, {Pancino}, {Recio-Blanco}, {Smiljanic}, {Carraro}, {Costado},
  {Damiani}, {Donati}, {Frasca}, {Jofr{\'e}}, {Lardo}, {de Laverny}, {Monaco},
  {Prisinzano}, {Sbordone}, {Sousa}, {Tautvai{\v{s}}ien{\.{e}}}, {Zaggia},
  {Zwitter}, {Delgado Mena}, {Chorniy}, {Martell}, {Silva Aguirre}, {Miglio},
  {Chiappini}, {Montalban}, {Morel}, \& {Valentini}}]{Casey2016}
{Casey}, A.~R., {Ruchti}, G., {Masseron}, T., {et~al.} 2016, \mnras, 461, 3336

\bibitem[{{Charbonnel} \& {Zahn}(2007)}]{Charbonnel2007}
{Charbonnel}, C. \& {Zahn}, J.~P. 2007, \aap, 467, L15

\bibitem[{{Cyburt} {et~al.}(2010){Cyburt}, {Amthor}, {Ferguson}, {Meisel},
  {Smith}, {Warren}, {Heger}, {Hoffman}, {Rauscher}, {Sakharuk}, {Schatz},
  {Thielemann}, \& {Wiescher}}]{Cyburt2010}
{Cyburt}, R.~H., {Amthor}, A.~M., {Ferguson}, R., {et~al.} 2010, \apjs, 189,
  240

\bibitem[{{Deepak} {et~al.}(2020){Deepak}, {Lambert}, \& {Reddy}}]{Deepak2020}
{Deepak}, {Lambert}, D.~L., \& {Reddy}, B.~E. 2020, \mnras, 494, 1348

\bibitem[{Deepak \& Reddy(2019)}]{Deepak2019}
Deepak \& Reddy, B.~E. 2019, Monthly Notices of the Royal Astronomical Society,
  484, 2000

\bibitem[{{Dupuis} {et~al.}(1992){Dupuis}, {Fontaine}, {Pelletier}, \&
  {Wesemael}}]{Dupuis1992}
{Dupuis}, J., {Fontaine}, G., {Pelletier}, C., \& {Wesemael}, F. 1992, \apjs,
  82, 505

\bibitem[{{Ferguson} {et~al.}(2005){Ferguson}, {Alexander}, {Allard}, {Barman},
  {Bodnarik}, {Hauschildt}, {Heffner-Wong}, \& {Tamanai}}]{Ferguson2005}
{Ferguson}, J.~W., {Alexander}, D.~R., {Allard}, F., {et~al.} 2005, \apj, 623,
  585

\bibitem[{{Gaia Collaboration}(2018)}]{Gaia2018}
{Gaia Collaboration}. 2018, VizieR Online Data Catalog, I/345

\bibitem[{{Gaia Collaboration} {et~al.}(2016){Gaia Collaboration}, {Prusti},
  {de Bruijne}, {Brown}, \& {Vallenari}}]{Gaia2016}
{Gaia Collaboration}, {Prusti}, T., {de Bruijne}, J.~H.~J., {Brown}, A.~G.~A.,
  \& {Vallenari}, A. 2016, \aap, 595, A1

\bibitem[{Gao {et~al.}(2019)Gao, Shi, Yan, Yan, Xiang, Zhou, Li, \&
  Zhao}]{Gao2019}
Gao, Q., Shi, J.-R., Yan, H.-L., {et~al.} 2019, The Astrophysical Journal
  Supplement Series, 245, 33

\bibitem[{Garaud(2018)}]{Garaud2018}
Garaud, P. 2018, Annual Review of Fluid Mechanics, 50, 275

\bibitem[{{Holanda} {et~al.}(2020){Holanda}, {Drake}, \&
  {Pereira}}]{Holanda2020}
{Holanda}, N., {Drake}, N.~A., \& {Pereira}, C.~B. 2020, \aj, 159, 9

\bibitem[{{Iben}(1967{\natexlab{a}})}]{Iben1967a}
{Iben}, Icko, J. 1967{\natexlab{a}}, \apj, 147, 650

\bibitem[{{Iben}(1967{\natexlab{b}})}]{Iben1967b}
{Iben}, Icko, J. 1967{\natexlab{b}}, \apj, 147, 624

\bibitem[{{Iglesias} \& {Rogers}(1993)}]{Iglesias1993}
{Iglesias}, C.~A. \& {Rogers}, F.~J. 1993, \apj, 412, 752

\bibitem[{{Iglesias} \& {Rogers}(1996)}]{Iglesias1996}
{Iglesias}, C.~A. \& {Rogers}, F.~J. 1996, \apj, 464, 943

\bibitem[{{Itoh} {et~al.}(1979){Itoh}, {Totsuji}, {Ichimaru}, \&
  {Dewitt}}]{Itoh1979}
{Itoh}, N., {Totsuji}, H., {Ichimaru}, S., \& {Dewitt}, H.~E. 1979, \apj, 234,
  1079

\bibitem[{{Kippenhahn} {et~al.}(1980){Kippenhahn}, {Ruschenplatt}, \&
  {Thomas}}]{Kippenhahn1980}
{Kippenhahn}, R., {Ruschenplatt}, G., \& {Thomas}, H.~C. 1980, \aap, 91, 175

\bibitem[{{Kirby} {et~al.}(2016){Kirby}, {Guhathakurta}, {Zhang}, {Hong},
  {Guo}, {Guo}, {Cohen}, \& {Cunha}}]{Kirby2016}
{Kirby}, E.~N., {Guhathakurta}, P., {Zhang}, A.~J., {et~al.} 2016, VizieR
  Online Data Catalog, J/ApJ/819/135

\bibitem[{{Kroupa} {et~al.}(1993){Kroupa}, {Tout}, \& {Gilmore}}]{Kroupa1993}
{Kroupa}, P., {Tout}, C.~A., \& {Gilmore}, G. 1993, \mnras, 262, 545

\bibitem[{{Kumar} {et~al.}(2020){Kumar}, {Reddy}, {Campbell}, {Maben}, {Zhao},
  \& {Ting}}]{Kumar2020}
{Kumar}, Y.~B., {Reddy}, B.~E., {Campbell}, S.~W., {et~al.} 2020, Nature
  Astronomy, 4, 1059

\bibitem[{{Kumar} {et~al.}(2011){Kumar}, {Reddy}, \& {Lambert}}]{Kumar2011}
{Kumar}, Y.~B., {Reddy}, B.~E., \& {Lambert}, D.~L. 2011, \apjl, 730, L12

\bibitem[{{Lattanzio} {et~al.}(2015){Lattanzio}, {Siess}, {Church}, {Angelou},
  {Stancliffe}, {Doherty}, {Stephen}, \& {Campbell}}]{Lattanzio2015}
{Lattanzio}, J.~C., {Siess}, L., {Church}, R.~P., {et~al.} 2015, \mnras, 446,
  2673

\bibitem[{{Li} {et~al.}(2018){Li}, {Aoki}, {Matsuno}, {Bharat Kumar}, {Shi},
  {Suda}, \& {Zhao}}]{Li2018}
{Li}, H., {Aoki}, W., {Matsuno}, T., {et~al.} 2018, \apjl, 852, L31

\bibitem[{{Lind} {et~al.}(2009){Lind}, {Primas}, {Charbonnel}, {Grundahl}, \&
  {Asplund}}]{Lind2009}
{Lind}, K., {Primas}, F., {Charbonnel}, C., {Grundahl}, F., \& {Asplund}, M.
  2009, \aap, 503, 545

\bibitem[{{Liu} {et~al.}(2014){Liu}, {Tan}, {Wang}, {Zhao}, {Sato}, {Takeda},
  \& {Li}}]{Liu2014}
{Liu}, Y.~J., {Tan}, K.~F., {Wang}, L., {et~al.} 2014, \apj, 785, 94

\bibitem[{{Lodders}(2019)}]{Lodders2019}
{Lodders}, K. 2019, in Nuclei in the Cosmos XV, Vol. 219, 165--170

\bibitem[{{L{\"u}} {et~al.}(2006){L{\"u}}, {Yungelson}, \& {Han}}]{L2006}
{L{\"u}}, G., {Yungelson}, L., \& {Han}, Z. 2006, \mnras, 372, 1389

\bibitem[{{L{\"u}} {et~al.}(2013){L{\"u}}, {Zhu}, \& {Podsiadlowski}}]{L2013}
{L{\"u}}, G., {Zhu}, C., \& {Podsiadlowski}, P. 2013, \apj, 768, 193

\bibitem[{{L{\"u}} {et~al.}(2020){L{\"u}}, {Zhu}, {Wang}, {Liu}, {Li}, {Xie},
  \& {Liu}}]{L2020}
{L{\"u}}, G., {Zhu}, C., {Wang}, Z., {et~al.} 2020, \apj, 890, 69

\bibitem[{{L{\"u}} {et~al.}(2009){L{\"u}}, {Zhu}, {Wang}, \& {Wang}}]{L2009}
{L{\"u}}, G., {Zhu}, C., {Wang}, Z., \& {Wang}, N. 2009, \mnras, 396, 1086

\bibitem[{{Martell} \& {Shetrone}(2013)}]{Martell2013}
{Martell}, S.~L. \& {Shetrone}, M.~D. 2013, \mnras, 430, 611

\bibitem[{{Martell} {et~al.}(2021){Martell}, {Simpson}, {Balasubramaniam},
  {Buder}, {Sharma}, {Hon}, {Stello}, {Ting}, {Asplund}, {Bland-Hawthorn}, {De
  Silva}, {Freeman}, {Hayden}, {Kos}, {Lewis}, {Lind}, {Zucker}, {Zwitter},
  {Campbell}, {{\v{C}}otar}, {Horner}, {Montet}, \& {Wittenmyer}}]{Martell2021}
{Martell}, S.~L., {Simpson}, J.~D., {Balasubramaniam}, A.~G., {et~al.} 2021,
  \mnras, 505, 5340

\bibitem[{{Monaco} {et~al.}(2011){Monaco}, {Villanova}, {Moni Bidin},
  {Carraro}, {Geisler}, {Bonifacio}, {Gonzalez}, {Zoccali}, \&
  {Jilkova}}]{Monaco2011}
{Monaco}, L., {Villanova}, S., {Moni Bidin}, C., {et~al.} 2011, \aap, 529, A90

\bibitem[{{Mori} {et~al.}(2021){Mori}, {Kusakabe}, {Balantekin}, {Kajino}, \&
  {Famiano}}]{Mori2020}
{Mori}, K., {Kusakabe}, M., {Balantekin}, A.~B., {Kajino}, T., \& {Famiano},
  M.~A. 2021, \mnras, 503, 2746

\bibitem[{{Paquette} {et~al.}(1986){Paquette}, {Pelletier}, {Fontaine}, \&
  {Michaud}}]{Paquette1986}
{Paquette}, C., {Pelletier}, C., {Fontaine}, G., \& {Michaud}, G. 1986, \apjs,
  61, 177

\bibitem[{{Paxton} {et~al.}(2011){Paxton}, {Bildsten}, {Dotter}, {Herwig},
  {Lesaffre}, \& {Timmes}}]{Paxton2011}
{Paxton}, B., {Bildsten}, L., {Dotter}, A., {et~al.} 2011, \apjs, 192, 3

\bibitem[{{Paxton} {et~al.}(2013){Paxton}, {Cantiello}, {Arras}, {Bildsten},
  {Brown}, {Dotter}, {Mankovich}, {Montgomery}, {Stello}, {Timmes}, \&
  {Townsend}}]{Paxton2013}
{Paxton}, B., {Cantiello}, M., {Arras}, P., {et~al.} 2013, \apjs, 208, 4

\bibitem[{{Paxton} {et~al.}(2015){Paxton}, {Marchant}, {Schwab}, {Bauer},
  {Bildsten}, {Cantiello}, {Dessart}, {Farmer}, {Hu}, {Langer}, {Townsend},
  {Townsley}, \& {Timmes}}]{Paxton2015}
{Paxton}, B., {Marchant}, P., {Schwab}, J., {et~al.} 2015, \apjs, 220, 15

\bibitem[{{Paxton} {et~al.}(2018){Paxton}, {Schwab}, {Bauer}, {Bildsten},
  {Blinnikov}, {Duffell}, {Farmer}, {Goldberg}, {Marchant}, {Sorokina},
  {Thoul}, {Townsend}, \& {Timmes}}]{Paxton2018}
{Paxton}, B., {Schwab}, J., {Bauer}, E.~B., {et~al.} 2018, \apjs, 234, 34

\bibitem[{{Paxton} {et~al.}(2019){Paxton}, {Smolec}, {Schwab}, {Gautschy},
  {Bildsten}, {Cantiello}, {Dotter}, {Farmer}, {Goldberg}, {Jermyn}, {Kanbur},
  {Marchant}, {Thoul}, {Townsend}, {Wolf}, {Zhang}, \& {Timmes}}]{Paxton2019}
{Paxton}, B., {Smolec}, R., {Schwab}, J., {et~al.} 2019, \apjs, 243, 10

\bibitem[{{Reimers}(1975)}]{Reimers1975}
{Reimers}, D. 1975, Memoires of the Societe Royale des Sciences de Liege, 8,
  369

\bibitem[{{Rogers} \& {Nayfonov}(2002)}]{Rogers2002}
{Rogers}, F.~J. \& {Nayfonov}, A. 2002, \apj, 576, 1064

\bibitem[{{Ruchti} {et~al.}(2011){Ruchti}, {Fulbright}, {Wyse}, {Gilmore},
  {Grebel}, {Bienaym{\'e}}, {Bland-Hawthorn}, {Freeman}, {Gibson}, {Munari},
  {Navarro}, {Parker}, {Reid}, {Seabroke}, {Siebert}, {Siviero}, {Steinmetz},
  {Watson}, {Williams}, \& {Zwitter}}]{Ruchti2011}
{Ruchti}, G.~R., {Fulbright}, J.~P., {Wyse}, R. F.~G., {et~al.} 2011, \apj,
  743, 107

\bibitem[{{Schwab}(2020)}]{Schwab2020}
{Schwab}, J. 2020, \apjl, 901, L18

\bibitem[{{Semenova} {et~al.}(2020){Semenova}, {Bergemann}, {Deal},
  {Serenelli}, {Hansen}, {Gallagher}, {Bayo}, {Bensby}, {Bragaglia}, {Carraro},
  {Morbidelli}, {Pancino}, \& {Smiljanic}}]{Semenova2020}
{Semenova}, E., {Bergemann}, M., {Deal}, M., {et~al.} 2020, \aap, 643, A164

\bibitem[{{Simonucci} {et~al.}(2013){Simonucci}, {Taioli}, {Palmerini}, \&
  {Busso}}]{Simonucci2013}
{Simonucci}, S., {Taioli}, S., {Palmerini}, S., \& {Busso}, M. 2013, \apj, 764,
  118

\bibitem[{{Singh} {et~al.}(2019){Singh}, {Reddy}, {Bharat Kumar}, \&
  {Antia}}]{Singh2019}
{Singh}, R., {Reddy}, B.~E., {Bharat Kumar}, Y., \& {Antia}, H.~M. 2019, \apjl,
  878, L21

\bibitem[{{Singh} {et~al.}(2021){Singh}, {Reddy}, {Campbell}, {Kumar}, \&
  {Vrard}}]{Singh2021}
{Singh}, R., {Reddy}, B.~E., {Campbell}, S.~W., {Kumar}, Y.~B., \& {Vrard}, M.
  2021, \apjl, 913, L4

\bibitem[{{Smiljanic} {et~al.}(2018){Smiljanic}, {Franciosini}, {Bragaglia},
  {Tautvai{\v{s}}ien{\.{e}}}, {Fu}, {Pancino}, {Adibekyan}, {Sousa}, {Randich},
  {Montalb{\'a}n}, {Pasquini}, {Magrini}, {Drazdauskas}, {Garc{\'\i}a},
  {Mathur}, {Mosser}, {R{\'e}gulo}, {de Assis Peralta}, {Hekker}, {Feuillet},
  {Valentini}, {Morel}, {Martell}, {Gilmore}, {Feltzing}, {Vallenari},
  {Bensby}, {Korn}, {Lanzafame}, {Recio-Blanco}, {Bayo}, {Carraro}, {Costado},
  {Frasca}, {Jofr{\'e}}, {Lardo}, {de Laverny}, {Lind}, {Masseron}, {Monaco},
  {Morbidelli}, {Prisinzano}, {Sbordone}, \& {Zaggia}}]{Smiljanic2018}
{Smiljanic}, R., {Franciosini}, E., {Bragaglia}, A., {et~al.} 2018, \aap, 617,
  A4

\bibitem[{{Stanton} \& {Murillo}(2016)}]{Stanton2016}
{Stanton}, L.~G. \& {Murillo}, M.~S. 2016, \pre, 93, 043203

\bibitem[{{Stephan} {et~al.}(2018){Stephan}, {Naoz}, \& {Gaudi}}]{Stephan2018}
{Stephan}, A.~P., {Naoz}, S., \& {Gaudi}, B.~S. 2018, \aj, 156, 128

\bibitem[{{Thomas}(1967)}]{Thomas1967}
{Thomas}, H.~C. 1967, in Late-Type Stars, ed. M.~{Hack}, 395

\bibitem[{{Thoul} {et~al.}(1994){Thoul}, {Bahcall}, \& {Loeb}}]{Thoul1994}
{Thoul}, A.~A., {Bahcall}, J.~N., \& {Loeb}, A. 1994, \apj, 421, 828

\bibitem[{{Timmes} \& {Swesty}(2000)}]{Timmes2000}
{Timmes}, F.~X. \& {Swesty}, F.~D. 2000, \apjs, 126, 501

\bibitem[{Ulrich(1972)}]{Ulrich1972}
Ulrich, R.~K. 1972, The Astrophysical Journal, 172, 165

\bibitem[{{Vescovi} {et~al.}(2019){Vescovi}, {Piersanti}, {Cristallo}, {Busso},
  {Vissani}, {Palmerini}, {Simonucci}, \& {Taioli}}]{Vescovi2019}
{Vescovi}, D., {Piersanti}, L., {Cristallo}, S., {et~al.} 2019, \aap, 623, A126

\bibitem[{{Wallerstein} \& {Sneden}(1982)}]{Wallerstein1982}
{Wallerstein}, G. \& {Sneden}, C. 1982, \apj, 255, 577

\bibitem[{{Yan} {et~al.}(2018){Yan}, {Shi}, {Zhou}, {Chen}, {Li}, {Zhang},
  {Bi}, {Wu}, {Li}, {Guo}, {Liu}, {Gao}, {Zhang}, {Zhou}, {Li}, \&
  {Zhao}}]{Yan2018}
{Yan}, H.-L., {Shi}, J.-R., {Zhou}, Y.-T., {et~al.} 2018, Nature Astronomy, 2,
  790

\bibitem[{{Yan} {et~al.}(2021){Yan}, {Zhou}, {Zhang}, {Li}, {Gao}, {Shi},
  {Zhao}, {Aoki}, {Matsuno}, {Li}, {Xu}, {Li}, {Wu}, {Jin}, {Mosser}, {Bi},
  {Fu}, {Pan}, {Suda}, {Liu}, {Zhao}, \& {Liang}}]{Yan2021}
{Yan}, H.-L., {Zhou}, Y.-T., {Zhang}, X., {et~al.} 2021, Nature Astronomy, 5,
  86

\bibitem[{{Yu} {et~al.}(2019){Yu}, {Li}, {Zhu}, {Wang}, {Liu}, {Guo}, {Han},
  {Chen}, \& {L{\"u}}}]{Yu2019}
{Yu}, J., {Li}, Z., {Zhu}, C., {et~al.} 2019, \apj, 885, 20

\bibitem[{{Yu} {et~al.}(2021){Yu}, {Zhang}, \& {L{\"u}}}]{Yu2021}
{Yu}, J., {Zhang}, X., \& {L{\"u}}, G. 2021, \mnras, 504, 2670

\bibitem[{{Zhang} {et~al.}(2021){Zhang}, {Shi}, {Yan}, {Li}, {Gao}, {Li},
  {Zhang}, {Liu}, {Bi}, {Zhao}, \& {Li}}]{Zhang2021}
{Zhang}, J., {Shi}, J.-R., {Yan}, H.-L., {et~al.} 2021, \apjl, 919, L3

\bibitem[{{Zhang} {et~al.}(2020){Zhang}, {Jeffery}, {Li}, \& {Bi}}]{Zhang2020}
{Zhang}, X., {Jeffery}, C.~S., {Li}, Y., \& {Bi}, S. 2020, \apj, 889, 33

\bibitem[{{Zhu} {et~al.}(2021){Zhu}, {Liu}, {Wang}, \& {L{\"u}}}]{Zhu2021}
{Zhu}, C., {Liu}, H., {Wang}, Z., \& {L{\"u}}, G. 2021, \aap, 654, A57

\end{thebibliography}
\end{document}